\documentclass[11pt,twoside]{article}

\usepackage{asp2006}
\usepackage{epsf}
\usepackage{psfig}

\begin{document}

\title{Post-Red Supergiants}   
\author{Ren\'e D. Oudmaijer$^1$, Ben Davies$^2$, Willem-Jan de Wit$^1$, Mitesh Patel$^3$}   
\affil{$^{1}$School of Physics $\&$ Astronomy, University of Leeds, Woodhouse Lane, Leeds LS2 9JT, UK \\
$^{2}$Center for Imaging Science, Rochester Institute of Technology, 54 Lomb Memorial Drive, Rochester NY 14623-5604, USA \\
$^{3}$Imperial College of Science, Technology and Medicine, Blackett Laboratory, Prince Consort Road, London SW7 2AZ, UK  }    

\begin{abstract} 
The yellow hypergiants are found in a stage between the massive Red
Supergiants and the Wolf-Rayet stars. This review\footnote{This is an
updated and slightly expanded version of a Keynote Talk given at ``Biggest, Baddest, Coolest
Stars'' (ASP Conf Series) eds. D. Luttermoser, B. Smith, and
R. Stencel} addresses current issues concerning the evolution of
massive stars, concentrating on the transitional post-Red Supergiant
phase.  Few yellow hypergiants are known and even fewer show direct
evidence for having evolved off the Red Supergiant branch. Indeed,
only two such rare objects with clear evidence for having gone through
of a previous mass losing phase are known, IRC +10420 and HD
179821. We will review their properties, discuss recent results
employing near-infrared interferometry, integral field spectroscopy
and polarimetry.  Finally, their real-time evolution is discussed.
\end{abstract}


\vspace*{-0.3cm}
\section{Introduction}  

A large variety of evolved massive stars are present in the upper
parts of the HR diagram. Amongst others, we find the hot, mass losing
Wolf-Rayet (WR) stars (Crowther, 2007), the unstable Luminous Blue
Variables (LBVs, Humphreys \& Davidson 1994), the enigmatic B[e]
supergiants (Zickgraf et al. 1986), the massive Red Supergiants (RSG,
Massey et al. 2008) and in between these phases, the Yellow Hypergiants
(YHGs, de Jager 1998). An HR diagram with LBVs, Yellow Hypergiants and
Red Supergiants is shown in Figure 1.

\begin{figure}
\plotfiddle{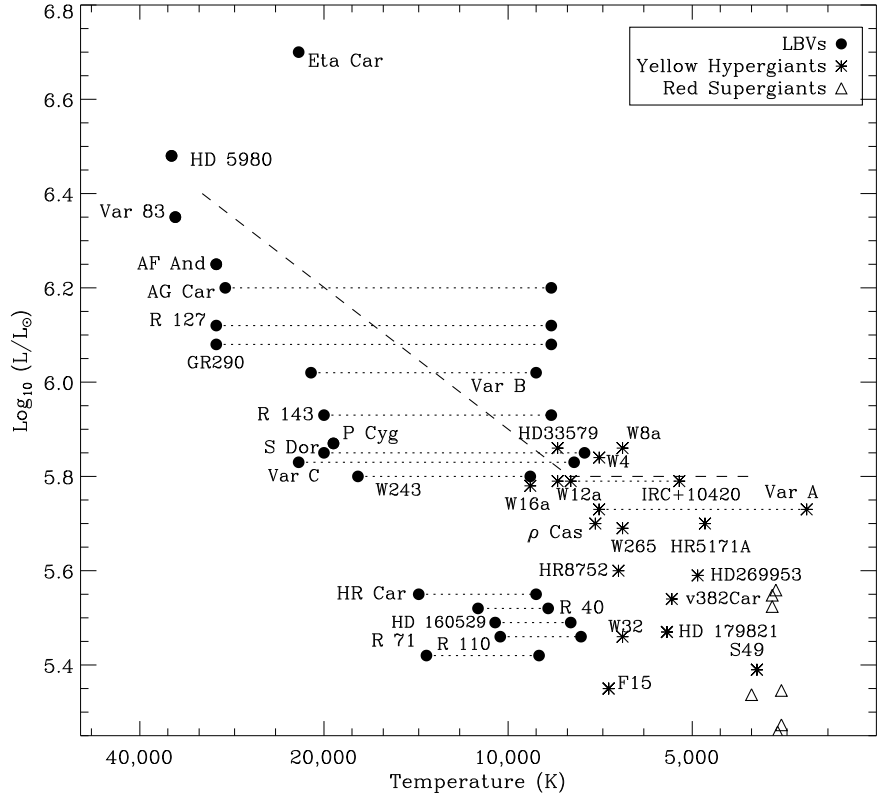}{7.cm}{90}{50}{50}{180}{0}
\caption{HR diagram displaying Luminous Blue variables (filled
circles, taken from Smith et al. 2004), yellow hypergiants (asterisks)
at their highest observed temperatures (de Jager 1998), and Red
Supergiants (triangles) from Levesque \& Massey (2005). Many Galactic
YHG data are taken from de Jager (1998), HD 179821 is from from Reddy
\& Hrivnak (1999). A large number of YHGs are now known from detailed
studies of massive Galactic clusters. F15 is reported by Figer et al.
(2006), Star 49 is presented by Davies et al. (2007a), and data for
W4, W8a, W12a, W16a, W32 and W265 in Westerlund 1 are taken from Clark
et al (2005).  The dashed line denotes the Humphreys-Davidson limit.}
\end{figure}

A commonly cited evolutionary scenario linking most of these classes
of object can be summarized as follows (e.g. Meynet \& Maeder
2003). After a massive star with a mass of more than about
50 $M_{\odot}$ moves off the main sequence it evolves straight to the
WR phase via the B supergiant and LBV phases. In special circumstances
it can then explode as a Supernova giving rise to a Gamma Ray Burst
(e.g. Meynet \& Maeder 2007). Lower mass stars with masses typically
larger than 10 $M_{\odot}$ evolve all the way from the blue to the red
to become a Red Supergiant, during which phase they can lose half
their mass via a strong, long lasting mass loss phase. When the star
evolves off the RSG branch it can evolve via the YHG phase to the LBV
phase to eventually become a WR star.  It should be mentioned however
that it is still not clear whether any LBVs have evolved off the RSG
(e.g. Lamers et al. 2001).  The situation proves even less settled for
the B[e] supergiants, it is possible that they constitute a rapidly
rotating sub-sample of the objects that evolve through the LBV phase.

In this paper, we focus on the evolutionary phase that immediately
follows the Red Supergiant stage, the post-RSG phase.  This term would
of course apply to all objects, including the WR stars, that have
visited the Red Supergiant branch at some point during their
evolution. For the purposes of this paper, we will confine ourselves
to the first stop in their journey from the RSG towards the blue. A
quick glance at the HR diagram brings us then immediately to the
Yellow Hypergiants. They are located on the red side of the LBVs and
the apparent empty region in the HR diagram where the temperatures are
between $\sim$7000 and 10 000 K. As outlined in Figure 1, the LBVs
vary in temperature between $\sim$9000 K up to the S Dor instability
strip, for which higher luminosities correspond to higher
temperatures. The YHGs are located on the red side of this region,
which has historically been referred to as the Yellow Void (e.g. de
Jager and Nieuwenhuijzen 1997), and have not been observed to cross it
bluewards. Given that the LBVs are found on the blue side of the
Yellow Void and also do not seem to cross it, while the larger number
of known LBVs and YHGs seem to fill in the ``void'' since the term was
introduced, it could be argued that ``Yellow Wall'' is perhaps a more
appropriate term.

Yellow hypergiant stars are not in a quiet phase of their
evolution. Observationally they have been found to exhibit explosive
events and to undergo multiple blue- to red movements and vice versa
in the HR diagram. After the stars evolve from the RSG branch and
reach temperatures around 7000 K, their envelopes become unstable and
a large mass loss ensues (e.g. Stothers \& Chin 2001). A large enough
mass loss rate can give rise to an optically thick wind, or in other
words, the central star is surrounded by a cool pseudo-photosphere.
After the eruption has occurred, the motion of the star in the HR
diagram reverses and the object now follows a redward loop returning
to the RSG branch. After the wind clears, the star will be on a
blueward loop again until it can ``bounce'' another time against the
Yellow Void.  This so-called ``bouncing against the Yellow Void'' has
been observed in several instances. A compelling observation is
presented by de Jager \& Nieuwenhuijzen in 1997. They observed HR 8752
and found that it had undergone at least two such ``bounces'' in the
preceding 30 years. In both cases, the bounce was associated with an
episode of increased mass loss. Just before the mass loss event, the
star had reached its maximum temperature, to decrease thereafter as
expected for a newly formed thick pseudo-photosphere.  The duration of
these episodes is around 10 years for HR 8752. This is shorter than
what has been observed for the object Var A in M33. Humphreys et
al. (2006), observed a spectral type variation for this object from F
to M and back to F again over a period of 45 years, and present
evidence that the high mass loss and subsequent clearing of the
pseudo-photosphere are responsible for the redward and blueward
motions in much the same manner as explained above. Similar events,
but at much smaller timescales have been observed for the famous
object $\rho$ Cas. Lobel et al. (2003) describe the latest outburst of
the star during which its temperature decreased from 7000 K to less
than 4000 K in a matter of a few hundred days, to increase back to
almost 6000 K in roughly the same amount of time.

Once a sufficient amount of mass has been ejected, and it is not yet
clear how much, the stellar envelope can survive the instabilities and
the star can evolve through the void to re-appear as an LBV. But this
may not apply to all YHGs as speculated by Smith et
al. (2004). Inspection of the HR diagram in Figure 1 reveals that
there may be a gap in luminosity around log$(L/L_{\odot})$ between
5.6..5.8 where there are apparently no LBV counterparts to the
YHGs. Smith et al. propose that for stars within this luminosity
range, their pseudo-photospheres will have even higher opacities due
to the bi-stability mechanism (e.g. Vink et al. 1999). Subsequently
they will not be visible as LBVs because the pseudo-photospheres
effectively result in a YHG spectrum for a longer time. Instead, the
objects that can cross the Yellow Void would only re-appear as the
much hotter Ofpe/WN9 stars or B[e] stars.

\begin{figure}

\plotfiddle{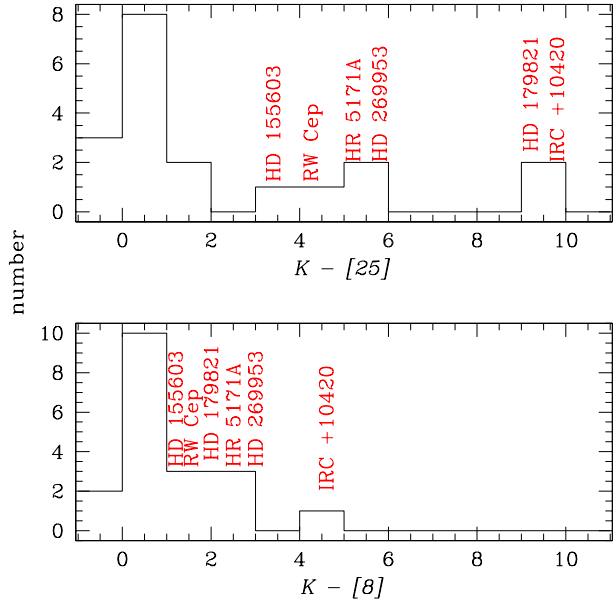}{8.cm}{0}{42}{42}{-150}{-50}
\caption{Infrared colours of the  K-G-F-A super- and hypergiants
listed in Table 2 of de Jager 1998. The upper panel shows the {\it
K $-$ [25]} colour in magnitudes (17 objects), that of the lower panel the
{\it K $-$ [8]} colour of 19 stars. See text for details.
}

\end{figure}

\section{Yellow Hypergiants as post-Red Supergiants}

In part because of their definition, the Yellow Hypergiants occupy a
rather limited area in the HR diagram.  It may come as a surprise then
that, apart from their status as evolved objects, their current
evolutionary phases are varied and for many individual cases their
precise status remains undetermined. Some objects have just left the
Main Sequence and are on their first, redward, move towards the Red
Supergiant Branch, some are on a blueward move, having just evolved
off the RSG branch, while others have left the Red Supergiant Branch
on a blueward track, but are returning on a redward loop.  With the
exception of the stars that are crossing the Yellow Void on a redward
track toward the RSG, all YHGs can be classified as post-Red
Supergiants. However, it is not trivial to determine the evolutionary
stage of a given YHG.

A key difference is that the post-Red Supergiants will have their
surface enriched by CNO processed material, often also traced by
their Na overabundance.  Unfortunately, not many detailed abundance
studies of YHGs have been performed. Klochkova et al. (1997) showed
for IRC +10420 that the star is N-rich, indicating dredge-up of
processed material. HD 179821 has been subject to 3 different
abundance studies with differing conclusions with respect to its
surface temperature. High temperatures, inconsistent with its G
spectral type have been proposed (Zacs et al. 1996), as have lower
temperatures (Reddy \& Hrivnak 1999; Th\'evenin et al. 2000). The main
issue appears to be methodological, namely how the Fe{\sc i} - Fe{\sc
ii} ionization balance can be properly reproduced, whilst at the same
time resulting in a G star spectrum. It would seem that a hydrogen
deficient photosphere could alleviate this problem, yet although hard
to demonstrate, this would also hint at a post-RSG nature of the
object.  Evidence for processed, enhanced sodium is reported for
$\rho$ Cas (El Eid \& Champagne 1995) and HR 5171A (Warren 1973),
while HD 33579 and HR 8752 were found to have solar abundances (de
Jager 1998 and Luck 1975 respectively). This has been taken as
evidence that HD 33579 is on a redward evolution, crossing the Yellow
Void for the first time. It should be stressed however that modern
abundance determinations using improved model atmospheres and high
resolution spectroscopy are lacking.

Another property of post-Red Supergiants is that they are surrounded
by the remnants of the prodigious mass loss during the RSG phase
itself.  Despite searches for the presence of circumstellar material,
not many YHGs have been found to exhibit evidence for a previous mass
loss episode. Some hypergiants have been reported in the literature to
have strong infrared excess emission due to circumstellar dust.  With
the new data from the Spitzer SAGE survey of the LMC (Meixner et
al. 2006) that have recently become available it is timely to
re-investigate the presence of infrared emission. We selected all
K-G-F-A supergiants from Table 2 in de Jager's 1998 review, but
excluded the well-known post-Asymptotic Giant Branch object AFGL 2688.
We then searched in the catalogues of the SAGE Spitzer survey, 2MASS,
the Gezari catalog, IRAS and MSX and plot their {\it K $-$ [8]} and {\it
K $-$ [25]} colours as histograms in Fig. 2. For two objects 8$\mu$m
fluxes were not available, and we used the IRAS 12$\mu$m flux
instead. Also, we treated MSX 21$\mu$m, Spitzer 24$\mu$m and IRAS
25$\mu$m fluxes equally. Of the 8 LMC supergiants listed in de Jager,
7 were detected by Spitzer.

The infrared bands probe the Rayleigh-Jeans tail of the stellar
photospheres and for a naked star the colours would be close to zero.
Clarke et al. (2005) showed that the {\it K $-$ [8]} and {\it K $-$
[25]} colours (with {\it [8]} and {\it [25]} denoting the magnitudes
at the respective wavelengths) are a powerful diagnostic of excess
emission due to circumstellar dust.  Whereas Clarke et al. could
identify excess objects with colours exceeding 0.4 magnitude, we have
to be more cautious here.  This is because the YHGs are variable and
not all photometry is taken simultaneously, and the colours may be
affected. This is illustrated by the fact that some stars would appear
to have negative excesses, which is unphysical.  Because of this, we
prefer to remain conservative and only assume the presence of excess
for the objects with colours larger than 2 magnitudes. This results in
4 objects with infrared excess.  HD 179821 and IRC +10420 stand out in
terms of their huge infrared excesses, and mass loss rates in excess
of 10$^{-4} \rm M_{\odot} yr^{-1}$ have been derived (Oudmaijer et
al. 1996, Hrivnak et al. 1989).  At longer wavelengths (not shown) the
excesses are even more pronounced. HR 5171A does not have much cool
dust, which would be indicative of a prolonged period of mass loss,
while for the Magellanic Cloud object HD 269953, longer wavelength
data beyond 25$\mu$m are lacking.

The infrared data are not alone in demonstrating the small number of
YHGs with evidence for a previous period of mass loss.  A very deep
HST imaging survey by Schuster et al. (2006) confirms this. They
observed the best-known YHGs, $\rho$ Cas, HR 8752 and HR 5171A, but did
not find any signs for extended emission. The only YHGs with evidence
for extended shells are HD 179821 and IRC +10420 (e.g. Kastner \&
Weintraub 1995). It would appear that these two stars are the only
ones with good observational evidence that they have evolved off the
Red Supergiant phase recently and are thus key to study the post-Red
Supergiant stage.

The lack of a detectable extended shell around the other objects
either implies that they are evolving from the Main Sequence to the
red for the first time, or that they have made one or more loops
between the RSG and the Yellow Void. We can not exclude the
possibility that the objects spend such a long time evolving slowly
towards the blue that the dust shell has long since dispersed into the
interstellar medium. This is not expected from evolutionary models
that can predict an RSG to Wolf-Rayet phase evolution of a few hundred
to a few thousand years (e.g. Garc\'{i}a-Segura et al. 1996), and many
more objects should be found than currently known.
In the remainder of this paper, we will continue with IRC +10420 and HD 179821.


\section{The post-Red Supergiants IRC +10420 and HD 179821}

As discussed above, the number of objects with evidence that they have
evolved off the Red Supergiant branch is sparse. Arguably, only
IRC +10420 is an undisputed post-Red Supergiant in the recent
literature. Its high luminosity (both spectroscopically and inferred
from its large distance), the large outflow velocity in CO (40
kms$^{-1}$) typical for a very luminous object and prodigious mass
loss are as expected for an object in the post-Red Supergiant stage
(e.g. Jones et al. 1993, Oudmaijer et al. 1996, Castro-Carrizo et
al. 2007).

The literature has been more ambivalent about the nature of HD 179821,
and we therefore take the opportunity to review the evidence that it
is a massive evolved star. In most papers up to the nineties it was
classed as a low to intermediate mass post-Asymptotic Giant Branch
(post-AGB) object. The post-AGB phase of evolution is qualitatively
almost identical to the post-RSG phase, as the star has evolved off
the AGB where it lost a lot of mass and evolves towards higher
temperatures. Many post-AGB stars display A-F-G supergiant spectra and
are surrounded by circumstellar material (see e.g. the reviews by van
Winckel, 2003, or Hrivnak, 2008).  However, several pieces of evidence
appear to indicate that HD 179821 is more likely to be a massive
object, and thus in the post-RSG phase instead.  To illustrate the
high luminosity of both objects, we show in Figure 3 the data and
calibration of the O{\sc i} $\lambda$7774 absorption system. This
triplet is a well known luminosity indicator and both IRC +10420 and
HD 179821 have strong absorptions, indicative of a very low surface
gravity. These two objects are even off the published scale and
clearly stand out from this sample of very bright supergiants. A
similar calibration was used by Clark et al. (2005), to determine the
luminosity of the YHGs in Westerlund I. Indeed, the method was one of
the primary methods to derive the distance to this cluster.

\begin{figure}
\plotfiddle{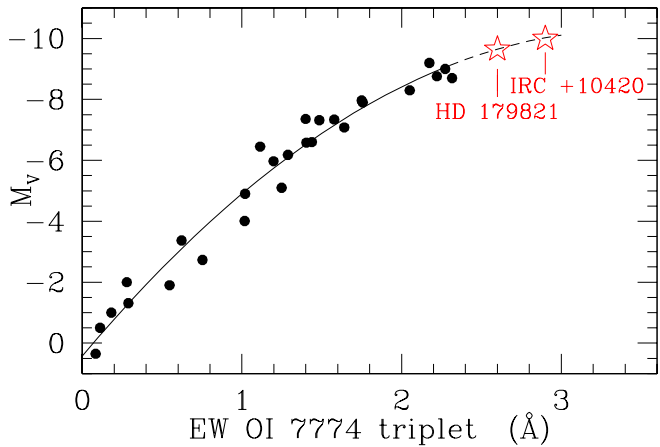}{6.5cm}{0}{45}{45}{-150}{-50}
\caption{The $M_V$ - $\lambda$7774 Equivalent Width calibration. The
data and fit are taken from Arellano Ferro et al. 2003. Both HD 179821
(data from Reddy \& Hrivnak 1999) and IRC +10420 (data from Oudmaijer
1998) are off this published scale and very luminous, indicating their
hypergiant nature.}
\end{figure}

Jura et al. (2001) compared both objects' space (Local Standard of
Rest, LSR) and expansion velocities of the mass outflows with
those of local low mass AGB stars and found them to be separated from
these lower mass objects both in space and expansion velocity. One can
derive a distance to the objects using their v$_{\rm LSR}$ velocities
and a model for the Galactic rotation curve. Jura et al. point out
that both objects are at a distance of several kpc and note that based
on the radial velocities alone, the luminosities are in excess of few
hundred thousand times solar.  The large CO outflow velocity of 30
kms$^{-1}$ found for HD 179821 is less high than for IRC +10420, but
very a-typical for lower mass AGB and post-AGB stars. To achieve such
high velocities a very luminous star is required to provide enough
radiation pressure to accelerate the flow (see Habing et al. 1994).

\begin{figure}
\plotfiddle{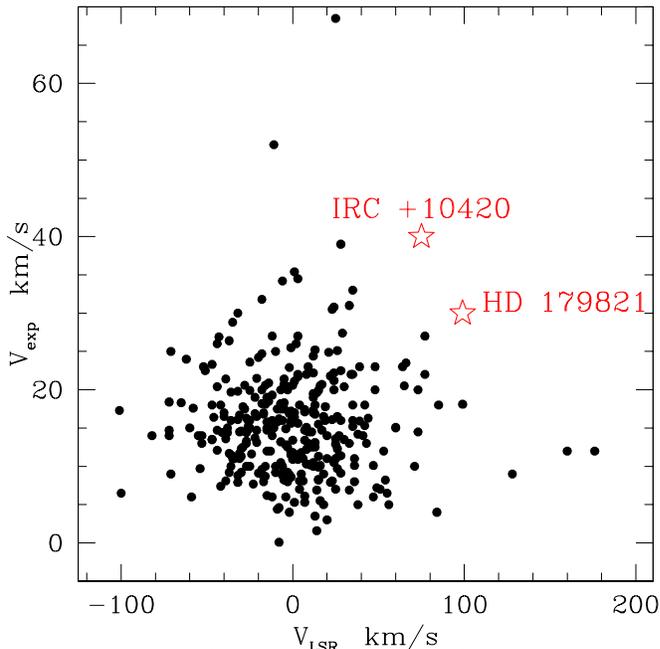}{7.5cm}{0}{45}{45}{-150}{-80}
\caption{CO expansion velocities and LSR velocities of a large sample
of evolved stars (Loup et al. 1993 - see text). Both post-Red
Supergiants stand out.}
\end{figure}

Inspired by the rather unusual, but informative, manner Jura et
al. (2001) plotted their data, we repeat the experiment using the
large compilation of CO observations of evolved objects due to Loup et
al. (1993).  We selected all stars with observations of the CO 1-0
line and available expansion velocities and LSR velocities. In case of
more than one entry, the first one in the list was taken. The
resulting sample contains 332 objects and their expansion velocities
are plotted against their space velocities in Fig. 4. Only the B-type
binary HD 101584 (Bakker et al. 1996) is not plotted as its high
outflow velocity of more than 100 kms$^{-1}$ would make the graph less
accessible.

We compute the average outflow velocity of this sample to be 16
kms$^{-1}$ with a scatter of 10 kms$^{-1}$. This is the typical
outflow velocity for low mass AGB stars and it can be seen that both
HD 179821 and IRC +10420 stand out both in space velocity and
expansion velocity. A number of objects have high outflow velocities
but are normal in terms of their LSR motion and vice versa. To check
whether there are objects with properties similar to post-RSG objects
in the sample, we studied these objects in a bit more detail. We find
26 objects with expansion velocities larger than 29 kms$^{-1}$
(arbitrarily chosen to be somewhat less than HD 179821's). Precisely
half of these are carbon-rich AGB stars, 5 are classified Planetary
Nebulae, 4 are oxygen-rich, including the well-known M supergiants VY
CMa and VX Sgr. The other two are the M stars V341 Car and
OH231.8+4.2. The remaining 4 objects include the aforementioned HD
101584 and the well-known Frosty Leo nebula (a K7III AGB/post-AGB
object with strong bi-polar jets, Castro-Carrizo et al. 2005). The
high CO velocities for these two objects reflect the presence of high
velocity bi-polar outflows, but do not necessarily represent the
outflow velocity of the preceding mass losing phase.  The final two
stars are HD 179821 and IRC +10420.

Since the objects in Loup's sample with known chemistry are equally
divided between carbon-rich and oxygen-rich objects, the preponderance
of C-rich AGB stars in the high velocity sub-sample may be a real
effect. As the stellar parameters of O- and C-rich AGB objects are
roughly the same, the reason for the higher velocities in some objects
is most likely to be found in the dust absorption properties of
carbon-rich dust. A higher absorption efficiency gives stronger
radiative driving and could result in some objects with large outflow
velocities.  A few objects with low outflow velocities, have large LSR
velocities. Although this is known (Feast et al. 2006), its origin is
unclear. It can be argued that the velocity dispersion of an evolved,
low mass population is large, and that some outliers can therefore
always be expected from a statistical point of view. It is also
possible that some of these objects are run-away stars after a
Supernova explosion from their companion.  As with Jura et al.'s 2001
graph, both post-Red Supergiants stand out in this sample, and there
is no indication of more such objects in Loup's catalog, although it
will be interesting to see whether similar objects have been observed
since.

The above two, very different, examples of investigating objects
illustrate the special place that both IRC +10420 and HD 179821 occupy
in different samples of object, and we conclude that HD 179821 is also
a massive star that has evolved off the post-Red Supergiant phase.

\subsection{Geometry of the circumstellar material}

Up til now, it is not clear which mechanism is responsible for the
shaping of the ring nebulae around WR stars, the aspherical
Supernovae ejecta and even the beamed Gamma Ray Bursts. A key question
is whether the shaping of the ejecta takes place during the Red
Supergiant phase, or thereafter (e.g. Dwarkadas \& Owocki 2002).

This is an area where the study of post-Red Supergiants can be
critical. The extended shells around both IRC +10420 and HD 179821
show considerable structure, but at the largest scales they are both,
to all intents and purposes, fairly spherically symmetric. This could
be seen already in the data of Kastner \& Weintraub (1995). The deep
images of Humphreys et al. (1997) show an almost annular appearance
for IRC +10420. Similarly, the HST data of Ueta et al. (2000) show a
predominately round shell for HD 179821. High resolution CO data by
Castro-Corrizo et al. (2007) confirm that the shells ejected during
the Red Supergiant phase are largely spherical.

There has been an interesting debate on the geometry of the H$\alpha$
line emitting region around IRC +10420. This line originates much
closer to the star than the dust and CO emission, and therefore traces
recent events. On the one hand, indirect evidence seemed to suggest it
deviates from spherical symmetry (e.g. Jones et al. 1993; Oudmaijer et
al. 1994), on the other hand, spherical symmetry has been proposed
(Humphreys et al. 2002). The latter authors used a novel and
innovative method to establish this. By taking spectra of the
reflection nebulosity around IRC +10420, they could effectively
observe the H$\alpha$ emission line from different
directions. This is because the dust particles located along the
rotation axis of the object can see a ``pole-on'' view of the
H$\alpha$ line emitting region, and scatter those photons to the
observer, the particles perpendicular to this axis would do the same,
but see, and scatter, an ``edge-on'' perspective.

\begin{figure}
\plotfiddle{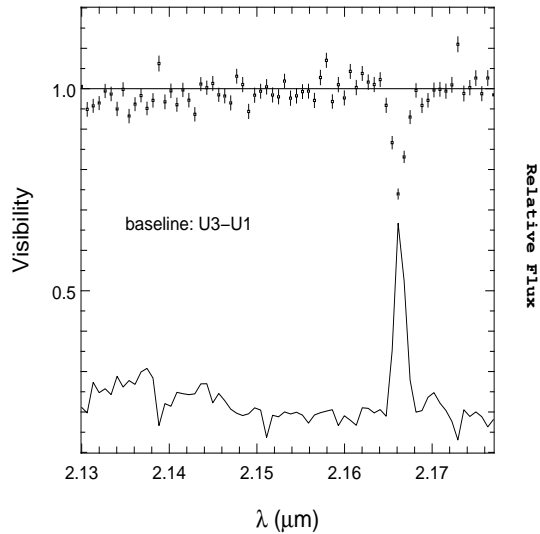}{7.5cm}{0}{45}{45}{-150}{-80}
\caption{AMBER/VLTI data of IRC +10420. The spectrum is centered at
Br$\gamma$ and has a resolution of 1200. The visibility resulting from
the longest baseline UT1-UT3 with a projected baseline of 69 m for the
observational set-up. The size of the line-emitting region is found to
be of order a few milli-arcseconds. From de Wit et al. 2008}
\end{figure}

Humphreys et al. (2002) employed the STIS instrument on board the HST
and observed IRC +10420 at two different slit positions. They found the
observed, reflected, H$\alpha$ profiles - taken as far as several
arcsec from the central star - to be similar and concluded that the
line forming region is spherical. An, unavoidable, limitation of their
observational setup is that only few viewing angles are
available. Below we present data that concern a full 360$^{\rm o}$
view of the H$\alpha$ emission. But let us first double check whether
the ionized region is indeed point-like in relation to the dust. This
is a tacit, but crucial, assumption when using the scattering method.

We obtained AMBER/VLTI interferometry of IRC +10420. This
near-infrared spectrograph uses the light combined from the 8.2 m ESO
Unit Telescopes UT1, UT2 and UT3 (e.g. Petrov et al. 2007). The data
were centered on the hydrogen recombination Br$\gamma$ emission
line. As the field of view of the instrument is 66 milli-arcseconds,
all the dusty emission that would be visible at the {\it K} band is
resolved out (see e.g. Bl\"ocker et al.'s 1999 visibility
calculations), and only the star and the Br$\gamma$ radiation - if
originating from a small region - is present. Our data at the longest
baseline are shown in Fig. 5 (from de Wit et al. 2008). The star is
unresolved at these wavelengths, the Br$\gamma$ line has a smaller
visibility and is clearly and unambiguously resolved. Preliminary
model fits to the data indicate that the diameter of the line emitting
region is 3.3 milli-arcseconds. This finding confirms that the
reflection method can be used, as the ionized region is located up to
a factor thousand closer to the star than the reflecting dust. 

\begin{figure}
\plotfiddle{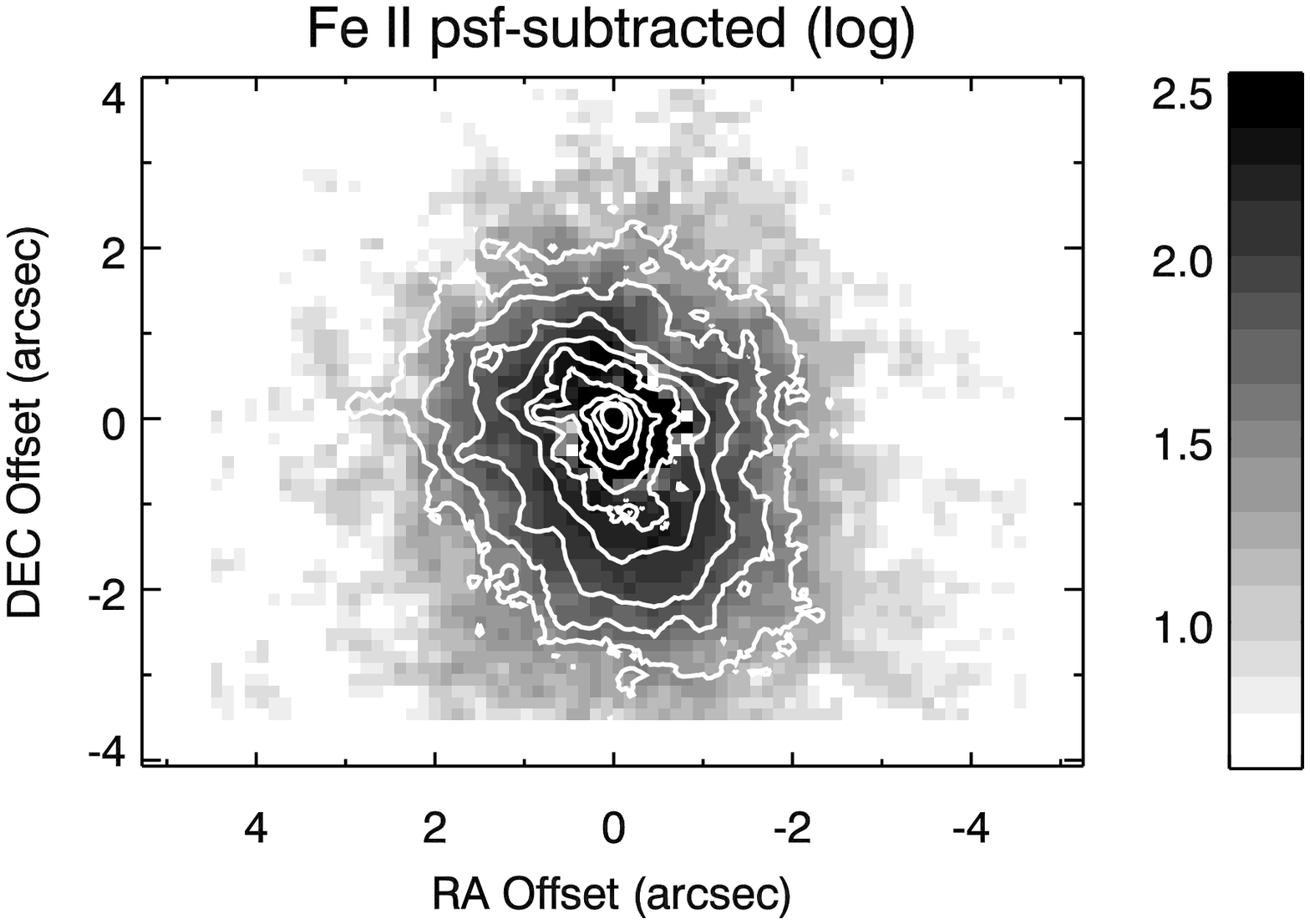}{2.5cm}{0}{35}{35}{-200}{-80}
\plotfiddle{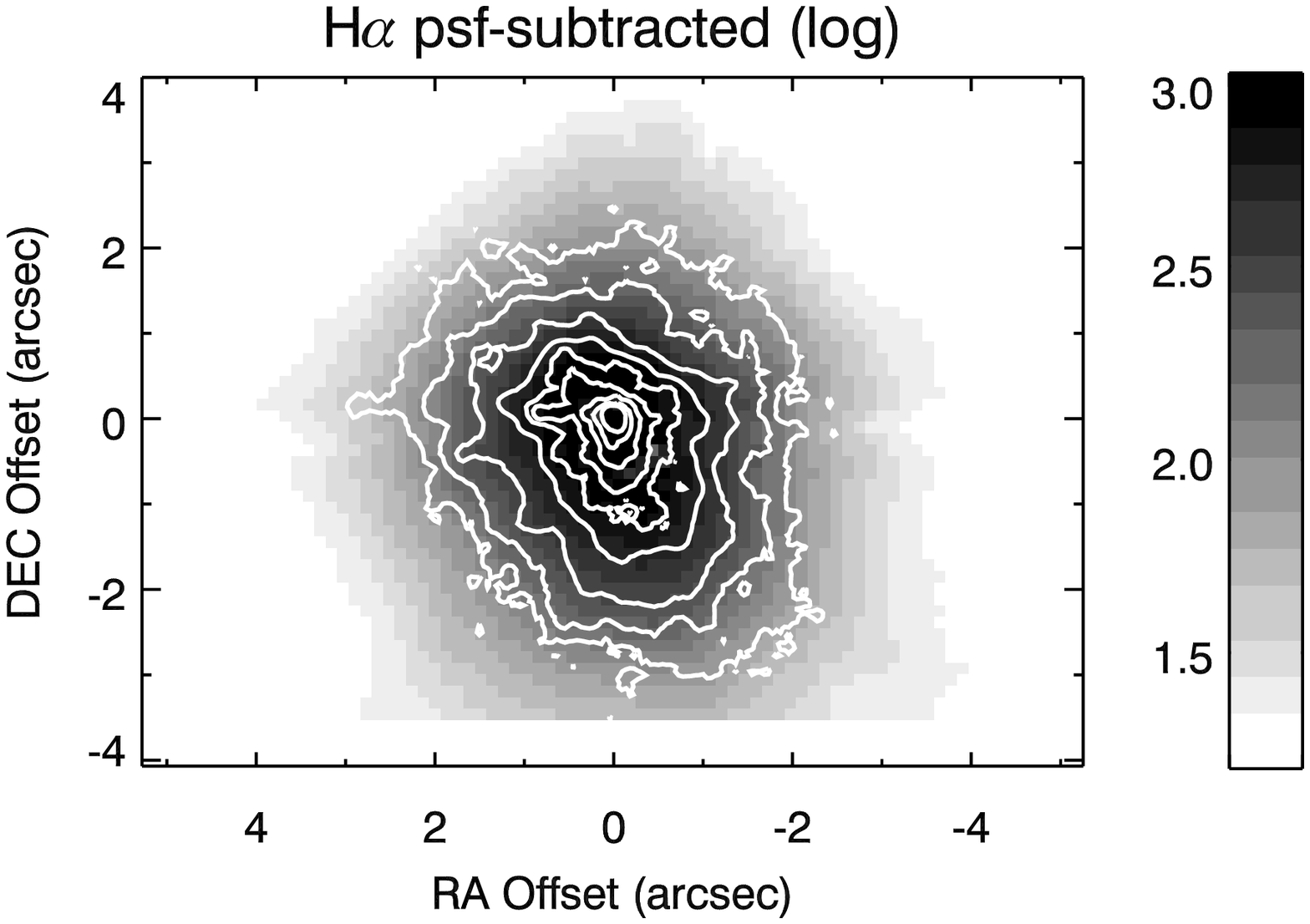}{2.5cm}{0}{36}{36}{0}{0}
\caption{Adaptive optics assisted IFU data of IRC +10420. The left
panel shows a continuum subtracted image of the H$\alpha$ emission
line. The right hand panel shows the Fe{\sc ii}$\lambda$ 6516
line. Both lines follow the reflection nebula, outlined by the
white contours, closely.}
\end{figure}

If the Br$\gamma$ emission is optically thin, then the observations do
not allow one to make any inferences about the shape of the line
forming region. This is because the baselines only sample a small
range of position angles on the sky (between 10$^{\rm o}$ and 30$^{\rm
o}$). However, for optically thick line emission, we can do this using
simple arguments.  The observed diameter, obtained along one position
angle, implies that the projected surface area is about ten times
larger than that of the star if the line forming region is
circular. If the Br$\gamma$ emission is optically thick, de Wit et
al. (2008) then demonstrate that the Br$\gamma$ emission line should
be an order of magnitude stronger than is actually observed. This
discrepancy can be explained if the geometry is not circular in
projection, but elongated. In this situation, the projected area of
the ionized region can be smaller and, therefore, a weaker emission
line is observed.  Future VLTI data are planned to improve the {\it
uv} coverage so that we can directly probe the geometry of the
circumstellar material. The future is exciting, with imaging
interferometers operating in the optical and near-infrared such as the
Magdalena Ridge Observatory Interferometer becoming in operation soon
(Creech-Eakman, 2008), we can begin to properly parametrize the winds
of evolved stars.

\begin{figure}
\plotfiddle{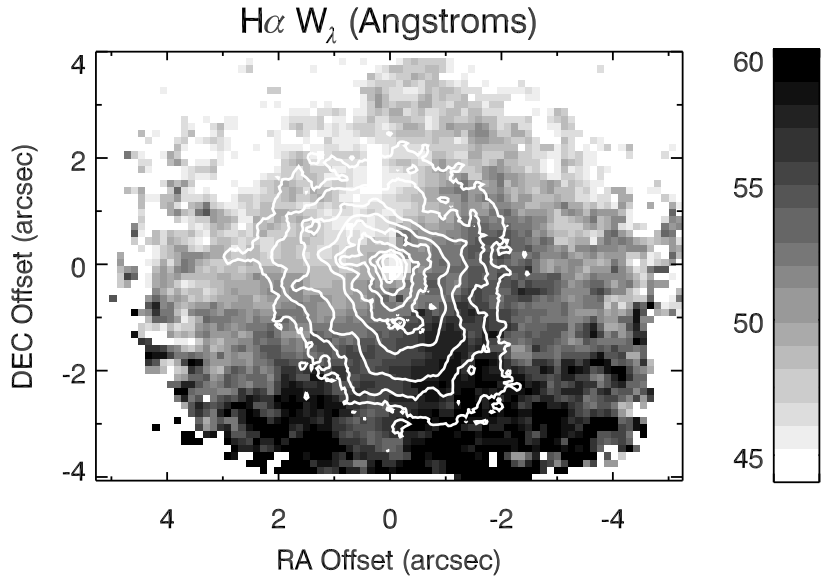}{7.5cm}{0}{60}{60}{-170}{0}
\caption{The H$\alpha$ emission Equivalent Width measured over the
reflection nebula. The variations indicate that the line-forming
region is not isotropic. }
\end{figure}

Returning to the reflected spectrum of the central star, we used
adaptive optics assisted integral field spectroscopy with the 4.2m
William Herschel Telescope (Davies et al. 2007b). The integral field
set-up covers the entire nebula and allows us to trace the reflected
line emission at more than only a handful of slit-positions.  The
spectra, with a spatial resolution of half an arcsecond covered the
H$\alpha$ line and a strong Fe{\sc ii} emission line at 6516$\rm
\AA$. The continuum subtracted images of both these lines are shown in
Fig. 6, overplotted is a contour plot of a (higher resolution) blue
HST image tracing the scattered light. The emission lines follow the
reflection nebula very closely and are an independent confirmation
that the line emission is reflected rather than  formed in situ.
The data do not have the spectral resolution to study the emission
line profiles in detail. Instead, the H$\alpha$ equivalent width is
shown in Fig. 7. The line EW is not constant, but exhibits a strong
variation over the nebulosity. The EW changes from around $-$60 $\rm
\AA$ in the south-west to less than $-50 \, \rm \AA$ in the
north-east.  As the equivalent width measures the line strength
compared to the stellar continuum (which is reflected equally in all
points), the change in EW indicates that the line emitting region is
{\it not} isotropic.  This result, which is in apparent contradiction
with the results from Humphreys et al. (2002), who did not find such
strong variations, can be readily explained. The slit positions
Humphreys et al.  employed turned out to probe a region where the
change in EW is minimal.

The axi-symmetry in the H$\alpha$ equivalent width is oriented
perpendicularly to the long axis of the nebula.  Although the current
result is strong evidence for an inhomogeneous H$\alpha$ emitting
region, the mechanism responsible for it can not be identified based
on these data. It will be interesting to follow-up this study at
higher spectral resolution to investigate the nature of the different
line profiles.

\subsection{Evolution of the objects}

Let us now turn to the evolution of both HD 179821 and IRC +10420.
Patel et al. (2008) collected optical and near-infrared photometric
data from the literature for both objects and the results are
presented in Fig. 8. The sources for the photometry are listed by
Patel et al. Only a few near-infrared data points have been reported
for HD 179821 over the last 20 years, which is particularly striking
as it is such a bright object. The properties of both stars in the
near-infrared photometry are very similar, the {\it J} and {\it K}
band magnitudes have become fainter over the past 40 and 20 years
respectively. The {\it V} band photometry shows
modest changes and is correlated with the near-infrared variability.

\begin{figure}
\plotfiddle{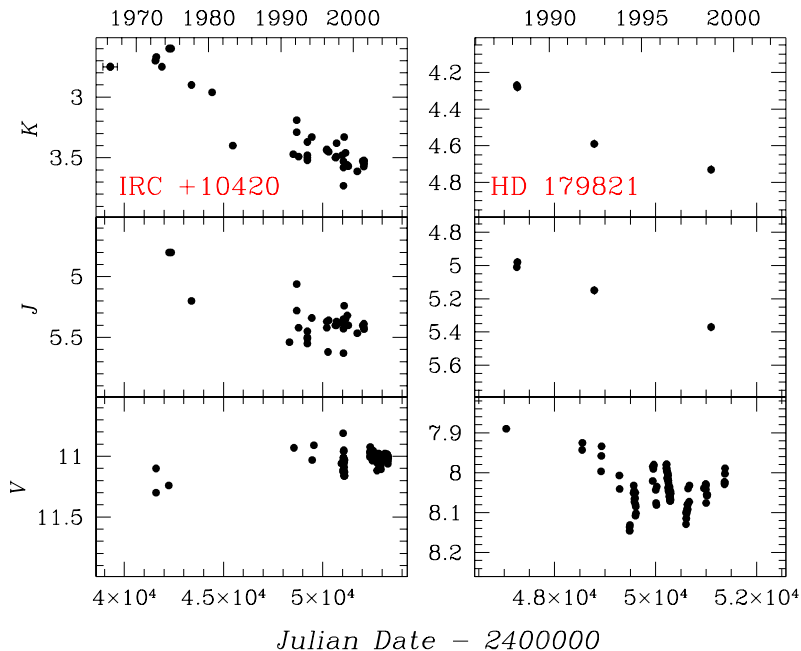}{7.9cm}{0}{54}{54}{-175}{-80}
\caption{Photometric history for both objects over the past
decades. Especially the changes in the {\it J} band, tracing the
stellar photosphere, indicate the temperature increase of the
stars. Data taken from Patel et al. (2008).}
\end{figure}

From model fits to the Spectral Energy Distributions, the {\it J} band
is known to trace the stellar photosphere (see Oudmaijer et al. 1996
and Hrivnak et al. 1989), so the changes do reflect variability of
the star itself.  Overall, the photometric changes can be explained by
a gradual increase in temperature of the objects. In the case of IRC
+10420 this has been confirmed spectroscopically. Klochkova et
al. (1997), in a study based on the data of Oudmaijer (1998), find
that the star must have a temperature consistent with an A supergiant
- as opposed to F8I derived from a spectrum 20 years earlier.  The
{\it J} and {\it V} band photometry have reached a steady state over
the past 10 years and although an increase in temperature is not
excluded by these data, it may well be that IRC +10420 has now hit the
``Yellow Void'' and will not evolve further to the blue. It is
tempting to speculate that a large outburst and a subsequent move to
the red as observed for $\rho$ Cas, HR 8752 and Var A is imminent. The
change in photometry for HD 179821 is a new result and needs
confirmation with spectroscopy. Although several spectroscopic studies
of HD 179821 have been published, as mentioned earlier, the abundance
studies unfortunately disagree on the temperature of the star, this is
due to different methodology rather than an evolutionary effect (see
for example the discussion by Th\'evenin et al. 2000).

To summarize, both stars appear to be evolving in real time. Whereas
IRC +10420 may have slowed down its evolution, HD 179821 is still
going strong and is evolving to the blue. As an aside, the added value of
the near-infrared {\it J} band proves to be a very useful diagnostic
to trace photospheric changes. The prospect of AAVSO observers being
able to obtain near-infrared photometry (Templeton, 2008) is very
exciting indeed.

\begin{figure}
\plotfiddle{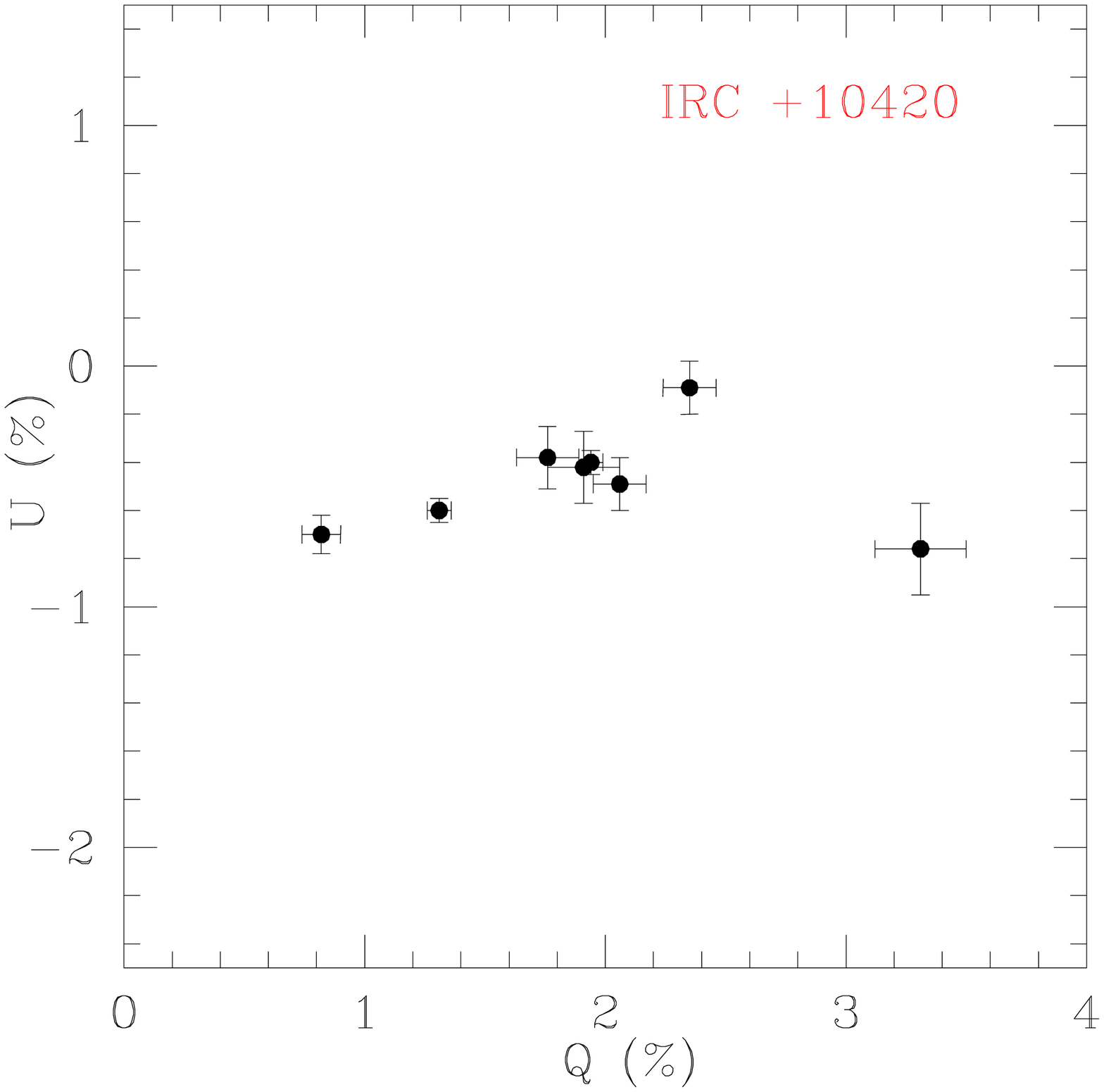}{2.5cm}{0}{30}{30}{-200}{-120}
\plotfiddle{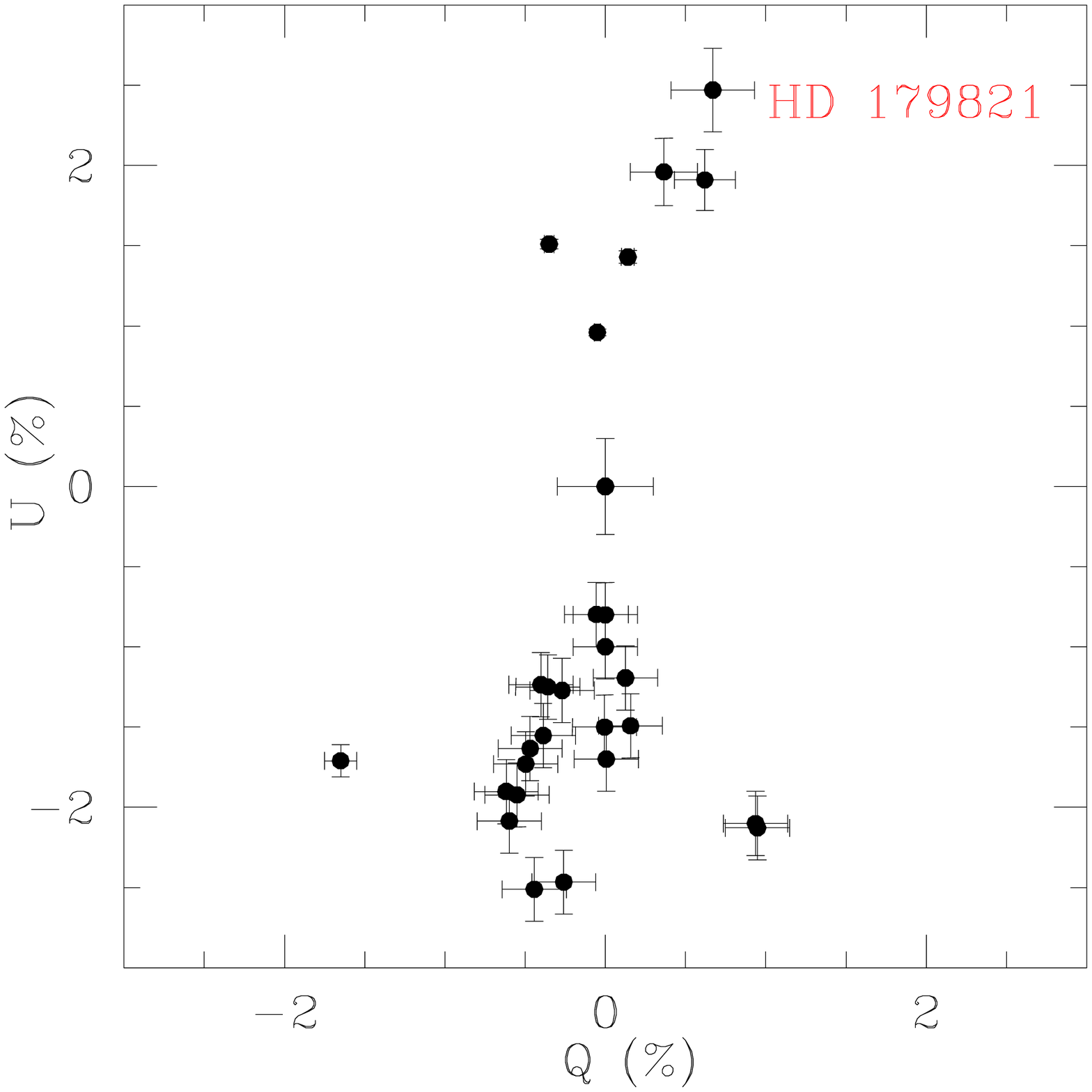}{2.5cm}{0}{31}{31}{0}{-40}
\caption{The Polarization Stokes QU parameters measured in the {\it R}
band of both stars taken over several decades (data from Patel et
al. 2008).}
\end{figure}

Patel et al. (2008) also collected polarization data from the
literature. They restricted themselves to {\it R} band data in order
to compare with their H$\alpha$ spectropolarimetric data. The results
represented in the Stokes QU vectors\footnote{The polarization $ P =
\sqrt{Q^2 + U^2}$} are shown in Fig. 9. The point by Trammell et
al. (1994) at (Q,U) = (1.7\%,$-$3.2\%) falls outside the plot
boundaries. As with the photometry, the polarization data span several
decades, and it is clear that both objects are highly variable in
polarization. Broadly speaking, the polarization smoothly changes over
time by several per cent, but the causes are different for both
objects. The polarization observed towards evolved stars can normally
be interpreted as due to the scattering of photospheric radiation by
either circumstellar free electrons or dust particles.  The material
needs to be distributed aspherically however. In the case of a
spherically symmetric shell, all polarization vectors cancel out and
no net polarization would be observed.  For other geometries, such as
a disk, not all vectors cancel out and a clear polarization signature
is observed.  If the scattering optical depth changes, due to
variations in mass loss or ionization for example, the net observed
polarization changes as well. This is precisely what we see for IRC
+10420, the polarization increase occurs in tandem with the increase
in H$\alpha$ equivalent width (Patel et al. 2008) and can be traced
back to a more heavily ionized, asymmetric ionized shell. For HD
179821, the situation is less clearcut. Based on the H$\alpha$
absorption, we can infer that the star has virtually no ionized gas
around it, and certainly not enough to induce a polarization of
several per cent. However as outlined above, the dust shell is
spherical as well, and it is very hard to reconcile the large
polarization, or its variability, with the classic picture of
polarization due to an aspherical circumstellar matter
distribution. Instead, Patel et al. propose that it is the star itself
which is an-isotropic. This scenario has been suggested, and is well
accepted, to explain the polarization properties of cool stars. For
example, Betelgeuse is observed to have large, bright convection cells
on the stellar surface. These illuminate the circumstellar material
an-isotropically and give rise to net polarization (Hayes 1984, see
also Ignace \& Henson 2008). HD 179821 is warmer than Betelgeuse, but
can be expected to have bright spots on the star as well. Any
variability can be due to the combined effect of its non-radial
pulsations (Le Coroller et al. 2003) and its stellar evolution
described above.

\section{Final Remarks}

The Yellow Hypergiants are in a short-lived phase between the Red
Supergiant stage and the end phases of massive star evolution. Yet,
although their numbers are small, their properties are varied. The
stars have been found to have very large mass loss rates at high
outflow velocities, move along various loops in the HR diagram,
bounce off a Yellow Void which they seem forbidden to cross, exhibit
eruptive moments of mass loss and generally are amongst the visually
brightest objects in any stellar population. They are key objects that
link the comparatively well understood Red Supergiants and the
Wolf-Rayet stars, not only in terms of evolutionary connections, but
also with regards to the development of the circumstellar structures.

Although the objects can be placed on an HR diagram, their unknown
distances remain a major limiting factor in deriving their physical
parameters.  Recent progress has made the study of the Yellow
Hypergiants or post-Red Supergiants in depth as a sample a reality. In
part this is due to better observations of the Magellanic Clouds
(e.g. Spitzer SAGE) which allowed us to detect circumstellar matter at
infrared wavelengths, while facilities such as the James Webb
Telescope will allow us to go one step further and determine the dust
properties and mass loss rates in great detail.  A major step forward
has been the progress in the detailed studies of massive Galactic
clusters. Work published only over the last few years has more than
doubled the number of known Galactic hypergiants (Clark et al. 2005,
Figer et al. 2006, Davies et al. 2007a, see Figure 1), and prospects
for the future are exciting, not only in terms of numbers of stars
discovered as well as in-depth studies of individual cluster members.

Further and prolonged monitoring, both spectroscopically, and
photopolarimetrically, of all objects, including the well-studied
Galactic ones, will continue to bring up new insights into the
evolutionary properties of the stars and the highest spatial
resolution, interferometric observations, will provide invaluable
information on the shaping of their nebulae.



\acknowledgements 
RDO is grateful for the support from the Leverhulme Trust
for awarding a Research Fellowship.  We thank Martin Groenewegen for
many useful discussions regarding this work.



\begin{thebibliography}{}
\bibitem[]{bla}
Arellano Ferro, A., Gridhar S., Rojo Arellano E. 2003, RvMAA 39, 3
\bibitem[]{bla}
Bakker E.J., Lamers H.J.G.L.M., Waters L.B.F.M., Waelkens C. 1996, A\&A 310, 861\bibitem[]{bla}
Bl\"ocker T., Balega Y., Hofmann K.-H. et al. 1999 A\&A 348, 805
\bibitem[]{bla}
Castro-Corrizo A., Quintana-Lacaci G., Bujarrabal V. et al. 2007, A\&A 465, 457
\bibitem[]{bla}
Castro-Carrizo A. Bujarrabal V., S\'anchez Contreras C. et al. 2005 A\&A 431, 979
\bibitem[]{bla}
Clark J.S., Negueruela I., Crowther P.A., Goodwin S.P. 2005, A\&A 434, 949
\bibitem[]{bla}
Clarke A.J., Oudmaijer R.D., Lumsden 2005, MNRAS 363, 1111
\bibitem[]{bla}
Creech-Eakman M. 2008, these proceedings
\bibitem[]{bla}
Crowther P.A. 2007, ARA\&A 45, 177
\bibitem[]{bla}
Davies B., Figer D.F., Kudritzki R.-P. et al. 2007a, ApJ 671, 781
\bibitem[]{bla}
Davies B., Oudmaijer R.D., Sahu K.C. 2007b ApJ 671, 2059
\bibitem[]{bla}
de Jager C. 1998, A\&AR 8, 145
\bibitem[]{bla}
de Jager C., Nieuwenhuijzen H. 1997, MNRAS 290, L50
\bibitem[]{bla}
de Wit W.J.M., Oudmaijer R.D., Groenewegen M.A.T. et al. 2008, A\&A in press, arXiv:0711.4975 [astro-ph] 
\bibitem[]{bla}
Dwarkadas V.V., Owocki S.P. 2002, ApJ 581, 1337
\bibitem[]{bla}
El Eid M.F., Champagne A.E. 1995, ApJ 451, 298
\bibitem[]{bla}
Feast M.W., Whitelock P.A., Menzies J.W. 2006, MNRAS 369, 791
\bibitem[]{bla}
Figer D.F., MacKenty J.W. Robberto M. et al. 2006, ApJ 643, 1166
\bibitem[]{bla}
Garc\'{i}a-Segura G., Langer N. Mac Low M.-M. 1996, A\&A 316, 133
\bibitem[]{bla}
Hayes D.P. 1984, ApJS 55, 179 
\bibitem[]{bla}
Hrivnak B.J. 2008, these proceedings
\bibitem[]{bla}
Hrivnak B.J., Kwok S., Volk K.M., 1989 ApJ 346, 265
\bibitem[]{bla}
Humphreys R.M., Davidson K. 1994, PASP 106, 1025
\bibitem[]{bla}
Humphreys R.M., Smith N., Davidson K. et al. 1997, AJ 114, 2778
\bibitem[]{bla}
Humphreys R.M., Davidson K., Smith N. 2002, AJ 124, 1026
\bibitem[]{bla}
Humphreys R.M., Jones T.J., Polomski E. et al. 2006, AJ 131, 2105
\bibitem[]{bla}
Ignace R., Henson G. 2008, these proceedings
\bibitem[]{bla}
Jones T.J., Humphreys R.M., Gehrz R.D. et al. 1993, ApJ 411, 323
\bibitem[]{bla}
Jura M., Velusamy T., Werner M.W., 2001, ApJ 556, 408
\bibitem[]{bla}
Kastner J.H., Weintraub D.A. 1995, ApJ 452, 883
\bibitem[]{bla}
Klochkova V.G., Chentsov E.L., Panchuk V.E. 1997, MNRAS 292, 19
\bibitem[]{bla}
Lamers H.J.G.L.M., Nota A., Panagia N. et al. 2001, ApJ 551, 764
\bibitem[]{bla}
Le Coroller H., L\`ebre A., Gillet D., Chapellier E. 2003, A\&A 400, 613
\bibitem[]{bla}
Levesque E.M., Massey P.2005, ApJ 628, 973
\bibitem[]{bla}
Lobel A., Dupree A.K., Stefanik R.P. et al. 2003, ApJ 583, 954
\bibitem[]{bla}
Loup C., Forveille T., Omont A., Paul J.F. 1993, A\&AS 99, 291
\bibitem[]{bla}
Luck R.E. 1975, ApJ 202, 743
\bibitem[]{bla}
Massey P., Plez B., Levesque E.M. et al.  2008, these proceedings arXiv:0708.2847 [astro-ph]
\bibitem[]{bla}
Meixner M., Gordon K.D., Indebetouw R. et al. 2006, AJ 132, 2268
\bibitem[]{bla}
Meynet G., Maeder A. 2003, A\&A 404, 975
\bibitem[]{bla}
Meynet G., Maeder A. 2007, A\&A 464, L11
\bibitem[]{bla}
Oudmaijer R.D. 1998, A\&AS 129, 541
\bibitem[]{bla}
Oudmaijer R.D., Geballe T.R., Waters L.B.F.M., Sahu K.C. 1994, A\&A 281, L33 
\bibitem[]{bla}
Oudmaijer R.D., Groenewegen M.A.T., Matthews H.E., Blommaert J.A.D.L., Sahu K.C. 1996, MNRAS 280, 1062
\bibitem[]{bla}
Patel M., Oudmaijer R.D., Vink J.S. et al. 2008, MNRAS in press, arXiv:0801.0878 [astro-ph]
\bibitem[]{bla}
Petrov R.G., Malbet F., Weigelt G., et al. 2007, A\&A, 464, 1
\bibitem[]{bla}
Reddy B.E., Hrivnak B.J. 1999, AJ 117, 1834
\bibitem[]{bla}
Schuster M.T., Humphreys R.M., Marengo M. 2006, AJ 131, 603
\bibitem[]{bla}
Smith N., Vink J.S., de Koter A. 2004, ApJ 615, 475
\bibitem[]{bla}
Stothers R.B., Chin C.-W. 2001, ApJ 560, 934
\bibitem[]{bla}
Templeton M. 2008, these proceedings
\bibitem[]{bla}
Th\'evenin F., Parthasarathy M., Jasniewicz G. 2000, A\&A 359, 138
\bibitem[]{bla}
Trammell S.R., Dinerstein H.L., Goodrich R.W. 1994, AJ 108, 984
\bibitem[]{bla}
Ueta T., Meixner M., Bobrowsky M. 2000, ApJ 528, 861
\bibitem[]{bla}
van Winckel H. 2003, ARA\&A 41, 391 
\bibitem[]{bla}
Vink J.S., de Koter A., Lamers H.J.G.L.M. 1999, A\&A 350, 181
\bibitem[]{bla}
Warren P.R. 1973, MNRAS 161, 427 
\bibitem[]{bla}
Zacs L., Klochkova V.G., Panchuk V.E., Spelmanis R. 1996, MNRAS 282, 1171
\bibitem[]{bla}
Zickgraf F.-J., Wolf B., Leitherer C., Appenzeller I., Stahl O. 1986, A\&A 163, 119
\end{thebibliography}
\end{document}